\documentclass[final,1p,times]{elsarticle}

\usepackage{graphicx}
\usepackage{dcolumn}
\usepackage{amsfonts}
\usepackage{amssymb,amsmath}

\journal{Computer Physics Communications}
\biboptions{sort&compress} 
 
\begin{document}

\begin{frontmatter}

\title{Eulerian Rotations of Deformed Nuclei for TDDFT Calculations}
\author[label1,label2]{D.A. Pigg}
\author[label1]{A.S. Umar}
\ead{umar@compsci.cas.vanderbilt.edu}
\author[label1]{V.E. Oberacker}
\address[label1]{Department of Physics and Astronomy, Vanderbilt University, Nashville, TN 37235, USA}
\address[label2]{Department of Natural Science and Mathematics, Lee University, Cleveland, TN 37320, USA}

\begin{abstract}
We discuss three practical methods for performing Eulerian rotations of Slater determinants in a three-dimensional
Cartesian geometry. In addition to the straightforward application of the active form of the
quantum mechanical rotation operator, we introduce two methods using a passive
position-space rotation followed by an active spin-space rotation, one after variation and the other
before variation. These methods can be used to initialize reactions involving deformed nuclei where
a particular alignment of the deformed nuclei with respect to the collision axis is desired.
We show that doing the rotation before the variation is the most efficient way of generating
such initial states.
\end{abstract}


\end{frontmatter}

\section{Introduction}

\begin{figure}[!htb]
\begin{center}
\includegraphics*[width=10.0cm]{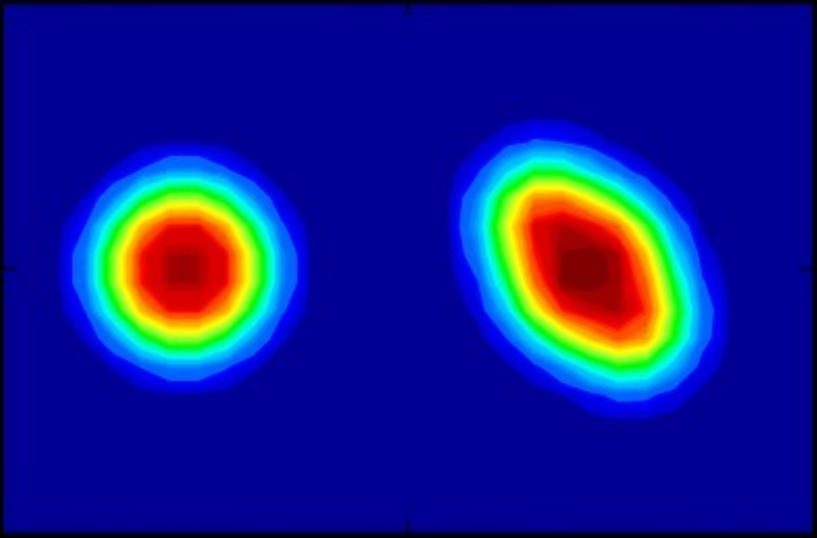}
\caption{Initial setup for the collision of $^{16}$O+$^{20}$Ne system, with
         the $^{20}$Ne nucleus at a $45^{o}$ angle with respect to the collision axis.
         In this calculation the nucleus $^{20}$Ne has a quadrupole deformation of $28.54$~fm$^2$.}
\label{fig1}
\end{center}
\end{figure}

It is generally acknowledged that the time-dependent density functional theory (TDDFT) provides a
useful foundation for a fully microscopic many-body theory of low-energy heavy-ion reactions
\cite{Ne82,Si12}. The TDDFT method is most widely known in nuclear physics in
the small amplitude domain, where it provides a useful description of collective states
\cite{UmOb05,SCh07,MR05,NY05,SC03}, and is based on the nuclear energy density functional theory,
which is a widely used approach for studying the static properties of nuclei~\cite{UNE1}.
TDDFT approach is also extensively used to study collisions of nuclei leading to fusion
and deep-inelastic heavy-ion collisions (for a recent review see Ref.~\cite{Si12}).
For collisions involving deformed nuclei in an unrestricted three-dimensional geometry 
the study of the collision for arbitrary alignment of the deformed nucleus with respect
to the collision axis is desirable. As an example in Fig.~\ref{fig1} we show the initialization of the
deformed $^{20}$Ne nucleus at a $45^{o}$ angle with the collision axis for the head-on
collision of the $^{16}$O+$^{20}$Ne system.

The most common choice for
a many-body wavefunction appropriate for DFT and TDDFT calculations is the Slater determinant.
In this manuscript we discuss three practical methods for performing Eulerian rotations of Slater
determinants in a three-dimensional Cartesian geometry.
In addition to the straightforward application of the active form of the
quantum mechanical rotation operator, we introduce two methods using a passive
position-space rotation followed by an active spin-space rotation, one after variation and the other
before variation. The {\it rotation before variation} approach can also be utilized to
generate exact initial states for time-dependent Hartree-Fock (TDHF) calculations involving
deformed nuclei at various orientations.
For increased numerical accuracy, the computations are done using basis-spline functions.
Basis-splines have been effectively used in HF~\cite{US91}, TDHF~\cite{UO06},
and HFB calculations~\cite{BO05,OU03,PS08}, however the procedures are easily adaptable to other
discretization methods.

Similarly, the description of a nucleus in terms of a self-consistent mean-field is one of the
most commonly used approaches in nuclear structure calculations~\cite{RS80,BR86}.
However, this approach
breaks many of the fundamental symmetries present in the nucleus, thus making it
desirable to restore some of these symmetries and consider the configuration mixing of
symmetry-restored mean-field states~\cite{BH03}. This allows the study of correlations in
addition to those present in the standard mean-field results (e.g. Pauli correlations).
The correlated states are usually generated by symmetry restoration via
the projection of angular momentum, parity, particle number (in case of pairing), 
and others~\cite{IM79,SG84,BH84,SG04,LD09}. Consequently, the techniques discussed here could also be
useful for these calculations.

\section{Active and Passive Rotations}

In the quantum theory of rotations, it is important to distinguish carefully between
active and passive transformations~\cite{S94}. In an \emph{active rotation} the wavefunction
of a quantum particle is rotated by an angle $\theta$, and the coordinate system is
left unchanged. In a \emph{passive rotation} the coordinate system is rotated by an angle $-\theta$,
and the wavefunction of a quantum particle is left unchanged.

In the following we consider an \emph{active rotation} of a single-particle spinor wavefunction (spin 1/2)
by an angle $\theta$ around an axis described by the unit vector \textbf{n}.
The rotation operator $R$ has the structure~\cite{BS79}
\begin{equation}
R(\theta,\mathbf{n}) = \exp \left [-i \frac{\theta}{\hbar} \mathbf{n} \boldsymbol{\cdot} \mathbf{J} \right ]\;  ,
\label{eq:R1}
\end{equation}
where \textbf{J} is the total angular momentum operator, consisting of orbital and
spin angular momenta
\begin{equation}
\mathbf{J} = \mathbf{L} + \mathbf{S} = -i \hbar ( \mathbf{r} \times\boldsymbol{\nabla} ) + \frac{\hbar}{2} \boldsymbol{\sigma} \; ,
\label{eq:J}
\end{equation}
and $\boldsymbol{\sigma}$ denotes the vector formed by the three Pauli matrices.
Instead of performing a rotation by an angle $\theta$ around the axis given by the vector
\textbf{n}, one usually parametrizes the rotation in terms of three
Euler angles $(\alpha,\beta,\gamma)$. First, one makes a rotation through an angle
$\gamma$ about the original z-axis, then a rotation through an angle
$\beta$ about the original y-axis, and finally a rotation through an angle
$\alpha$ about the original z-axis. Thus the operator for an active rotation has the
structure
\begin{equation}
R(\Omega) = R(\alpha,\beta,\gamma) = e^{-i \alpha J_z / \hbar} e^{-i \beta J_y / \hbar}
e^{-i \gamma J_z / \hbar}\; .
\label{eq:R2}
\end{equation}
We insert Eq.~(\ref{eq:J}) into Eq.~(\ref{eq:R2}) and observe that rotations in
position space, generated by \textbf{L}, and rotations in spin space, generated by
\textbf{S}, commute. This leads to
\begin{equation}
R(\Omega) = R_s(\Omega) \ R_r(\Omega)\; ,
\label{eq:R3}
\end{equation}
with the spin space rotation operator
\begin{equation}
R_s(\Omega) = e^{-i \alpha \sigma_z / 2} e^{-i \beta \sigma_y / 2}
e^{-i \gamma \sigma_z / 2}\; ,
\label{eq:Rs1}
\end{equation}
and the position space rotation operator
\begin{equation}
R_r(\Omega) = e^{-i \alpha L_z / \hbar} e^{-i \beta L_y / \hbar}
e^{-i \gamma L_z / \hbar} \; .
\label{eq:Rr1}
\end{equation}
We use the general expression for the spin rotation operator~\cite{S94}
\begin{equation}
R_s(\theta,\mathbf{n}) = e^{-i \theta \mathbf{n} \boldsymbol{\cdot} \boldsymbol{\sigma} / 2} = \mathrm{cos}\frac{\theta}{2}\; \mathbf{I}_s
- i \mathrm{sin}\frac{\theta}{2}\; \mathbf{n} \boldsymbol{\cdot} \boldsymbol{\sigma}\; ,
\label{eq:Rs2}
\end{equation}
where $\mathbf{I}_s$ denotes the $2 \times 2$ unit matrix in spin space. Using this
expression to calculate the operators in Eq.~(\ref{eq:Rs1}), we obtain the
desired form of the spin rotation operator in terms of the
Euler angles
\begin{displaymath}
\label{eq:Rs3}
R_s(\Omega) =
\begin{pmatrix}
e^{-i \frac{\alpha}{2}} & 0 \\
\ & \ \\
0 & e^{+i \frac{\alpha}{2}}
\end{pmatrix}
\begin{pmatrix}
\mathrm{cos}\frac{\beta}{2}  & -\mathrm{sin}\frac{\beta}{2}\\
\ & \ \\
\mathrm{sin}\frac{\beta}{2}   & \mathrm{cos}\frac{\beta}{2}
\end{pmatrix}
\begin{pmatrix}
 e^{-i \frac{\gamma}{2}} & 0 \\
\ & \ \\
0 & e^{+i \frac{\gamma}{2}}
\end{pmatrix}
\end{displaymath}
or explicitly
\begin{equation}
R_s(\Omega) =
\begin{pmatrix}
e^{-i \frac{\alpha}{2}}\mathrm{cos}\frac{\beta}{2}e^{-i \frac{\gamma}{2}} &  -e^{-i \frac{\alpha}{2}} \mathrm{sin}\frac{\beta}{2}
 e^{+i \frac{\gamma}{2}}\\
\ & \ \\
e^{+i \frac{\alpha}{2}} \mathrm{sin}\frac{\beta}{2} e^{-i\frac{\gamma}{2}} &   e^{+i \frac{\alpha}{2}} \mathrm{cos}\frac{\beta}{2}
e^{+i \frac{\gamma}{2}}
\end{pmatrix}\; .
\label{eq:Rs4}
\end{equation}
We now apply the rotation operator,  Eq.~(\ref{eq:R3}), to a given single-particle spinor
wave function $\boldsymbol{\psi}$ to obtain the rotated spinor wave function $\boldsymbol{\psi}'$
\begin{equation}
\label{eq:R6}
\left ( \begin{array}{c}
         \psi'_1(\mathbf{r})  \\
         \psi'_2(\mathbf{r})
        \end{array} \right ) = \ R_s(\Omega) \ R_r(\Omega)
\left ( \begin{array}{c}
         \psi_1(\mathbf{r})  \\
         \psi_2(\mathbf{r})
        \end{array} \right ) \ .
\end{equation}

The expression given in Eq.~(\ref{eq:R6}) is the most common approach for
rotating Slater determinants by rotating single-particle spinors.
However, one can formulate an alternate approach where the single-particle
spinors are unchanged but the coordinate system is rotated instead. This
corresponds to a passive rotation in position space, and it
can be expressed in the form~\cite{BH82}
\begin{equation}
\label{eq:R7}
R_r(\Omega)
\left ( \begin{array}{c}
         \psi_1(\mathbf{r})  \\
         \psi_2(\mathbf{r})
        \end{array} \right ) =
\left ( \begin{array}{c}
         \psi_1(\mathbf{r}')  \\
         \psi_2(\mathbf{r}')
        \end{array} \right ) \; ,
\end{equation}
with
\begin{equation}
\mathbf{r}' = R_r^{-1}(\Omega) \ \mathbf{r} \; .
\label{eq:R8}
\end{equation}
The operator $R_r^{-1}(\Omega)$ denotes the inverse of the operator
for an active rotation in position space or, equivalently, the operator for a passive rotation
of the coordinate system. This operator is given by~\cite{BH82}
\begin{eqnarray}
\label{eq:R9}
R_r^{-1}(\Omega) &=&
\begin{pmatrix}
\mathrm{cos}\gamma & \mathrm{sin}\gamma & 0 \\
-\mathrm{sin}\gamma & \mathrm{cos}\gamma & 0 \\
 0 & 0 & 1
\end{pmatrix}
\begin{pmatrix}
\mathrm{cos}\beta & 0 & -\mathrm{sin}\beta \\
0 & 1 & 0 \\
\mathrm{sin}\beta & 0 & \mathrm{cos}\beta
\end{pmatrix}
\nonumber \\
&\times &
\begin{pmatrix}
 \mathrm{cos}\alpha & \mathrm{sin}\alpha & 0 \\
 -\mathrm{sin}\alpha & \mathrm{cos}\alpha & 0 \\
0 & 0 & 1
\end{pmatrix}\; ,
\end{eqnarray}
or, explicitly~\cite{AW95}
\begin{equation}
R_r^{-1}(\Omega) =
\begin{pmatrix}
\mathrm{cos}\gamma \mathrm{cos}\beta \mathrm{cos}\alpha -\mathrm{sin}\gamma \mathrm{sin}\alpha
                                   & \mathrm{cos}\gamma \mathrm{cos}\beta \mathrm{sin}\alpha +\mathrm{sin}\gamma \mathrm{cos}\alpha
                                   & -\mathrm{cos}\gamma \mathrm{sin}\beta \\
\\
                                   -\mathrm{sin}\gamma \mathrm{cos}\beta \mathrm{cos}\alpha -\mathrm{cos}\gamma \mathrm{sin}\alpha
                                   & -\mathrm{sin}\gamma \mathrm{cos}\beta \mathrm{sin}\alpha +\mathrm{cos}\gamma \mathrm{cos}\alpha
                                   & \mathrm{sin}\gamma \mathrm{sin}\beta \\
\\
                                   \mathrm{sin}\beta \mathrm{cos}\alpha
                                   & \mathrm{sin}\beta \mathrm{sin}\alpha
                                   & \mathrm{cos}\beta
\end{pmatrix}\; .
\end{equation}

In summary, we can compute the rotated single-particle spinor wave function $\boldsymbol{\psi}'$ via
the operations
\begin{equation}
\left ( \begin{array}{c}
         \psi'_1(\mathbf{r})  \\
         \psi'_2(\mathbf{r})
        \end{array} \right ) = \ R_s(\Omega)
\left ( \begin{array}{c}
         \psi_1(\mathbf{r}')  \\
         \psi_2(\mathbf{r}')
        \end{array} \right )\; ,
\end{equation}
where the spin rotation matrix is given in Eq.~(\ref{eq:Rs4}) and the
rotated position vector is given in terms of Eq.~(\ref{eq:R8}) and
Eq.~(\ref{eq:R9}).

Generalization of the above transformations to a many-body Slater determinant
is straightforward due to the one-body nature of the involved operators. One
finds that the rotation operator acting on a Slater determinant
$\Psi$ for an $A$-nucleon system
results in a new Slater determinant $\Psi'$ which consists of the rotated
single-particle spinor wave functions given in Eq.~(14).
We also note here that if the calculations are done in a basis comprised of analytic
functions the implementation of such rotations will become straightforward.

\section{Application and Results}

In the sections below, we discuss three methods that can be used to perform
an Eulerian rotation of a Slater determinant and provide relative accuracies for
the methods. The first two methods deal with rotations performed after
the variational solution of the Hartree-Fock (HF) equations. The other method
involves the initialization of the HF calculations with a rotated initial guess
and the subsequent imaginary-time minimization~\cite{BSU}.
A detailed description of our three-dimensional, unrestricted HF code
is given in Refs~\cite{US91,UO06}.
For the effective interaction we have used the Skyrme SLy4 force~\cite{CB98}.
Since all the calculations are done with no geometrical symmetry restrictions, the
single-particle spinors carry only state indices that count the state number
and isospin quantum number.
For coordinate-space lattice solutions of the mean-field equations, a
very accurate representation of derivative operators and functions is necessary for
performing accurate rotations. This may be understood by viewing the rotation as
an interpolation onto a rotated coordinate system, whereby the mapping of the
original lattice points to the new points results in the interpolation of the
single-particle states, which requires high accuracy to preserve the integrity
of the system. Basis-spline interpolation~\cite{UW91} provides one such accurate approach, and
we use it in our HF calculations. For all of our computations, we have tried
to maintain the numerical details consistent with performing realistic calculations.
For test purposes we used a deformed $^{20}$Ne nucleus on a $30^3$ lattice with
a box size of $(-10, +10)$~fm in each direction.

\subsection{Rotation After Variation}

There are powerful methods for performing rotations of HF Slater determinants for
nuclear structure calculations that improve the mean-field results by using
correlated states~\cite{BH84,BH86,BH08}. Here, we choose two methods that were
readily available in our code.

\subsubsection{\label{sec:direct}Active Rotation via Eq.~\protect\ref{eq:R2}}
This approach is the commonly-used method for implementing the action of the rotation
operator onto a Slater determinant. The rotation is done after the minimum-energy solution to
the HF equations is achieved.
For Euler angles $(\alpha, \beta, \gamma)$, we apply Eq.~(\ref{eq:R2}) to each of the single-particle states
directly by breaking up the action of the exponential operators into small angular steps $(\Delta\alpha, \Delta\beta, \Delta\gamma)$,
where each step is carried out by the expansion of the exponential as a Taylor series. This yields, for example,
for a rotation by angle $\alpha$ about $\hat{z}$, the recurrence relation~\cite{SH06}
\begin{table}[!htb]
\caption{Rotation for Euler angle $\beta=45^{\circ}$ as a function of basis-spline order $M$
for the rotation method of Section~\ref{sec:direct}. The exact values for
the tabulated quantities are: $E_B=-157.26$~MeV, $r_{rms}=2.92$~fm, $\epsilon_{n}=-13.24$~MeV,
and $\epsilon_{p}=-9.35$~MeV. }
\label{tb:M3O} 
\centering
\begin{tabular}{lrrrr}
\hline\hline
& M=3 & M=5 & M=7 & M=9 \\
\hline
$E_B$          & -165.69 & -157.37 & -157.26 & -157.26 \\
$r_{rms}$      &    2.95 &    2.92 &    2.92 &    2.92 \\
$\epsilon_{n}$ &  -12.84 &  -13.21 &  -13.24 &  -13.24 \\
$\epsilon_{p}$ &   -9.01 &   -9.33 &   -9.34 &   -9.35 \\
\hline
\end{tabular}
\end{table}
\begin{equation}
|\psi_{\lambda}^{\alpha + \Delta \alpha}> \cong \sum_{n=0}^{N} \frac{(-i \Delta \alpha \hat{J}_z/\hbar)^n}{n!} |\psi_{\lambda}^{\alpha}>\; .
\label{eq:Shim}
\end{equation}
The choice of $N=4$ seems adequate for angular steps on the order of $\Delta \alpha=2\pi/360$.
The operations given in Eq.~(\ref{eq:Shim}) require the action of various derivative operators
onto the single-particle states. For example, for the case above,
$\hat{J}_z$ has the differential form
\begin{equation}
\hat{J}_z = i\hbar(y\partial_x-x\partial_y)\hat{I} + \frac{\hbar}{2}\hat{\sigma}_z\; .
\end{equation}
Similar expressions are for the rotation of $\beta$ about $\hat{y}$ and that of $\gamma$ about $\hat{z}$.
In the Basis-Spline Collocation Method (BSCM)~\cite{UW91}, the derivative operators for each
coordinate become matrices represented on the lattice and operate on the discretized single-particle
states. Thus we are able to compute the above expansion far more accurately than can be done with finite-difference
discretization.

In testing this method, we have found that the maximum error occurs for $45^{\circ}$ rotations
for each of the Euler angles and is worst for angle $\beta$. The reason is that
this orientation requires the most significant interpolation accuracy relative to
the original code frame
(although we are not doing interpolation here, the discretization of a differential equation
can be viewed as an interpolation problem) . In Table~\ref{tb:M3O} we show the results for the $45^{\circ}$
$\beta$ rotation as a function of the basis-spline order $M$. Tabulated are the
total HF binding energy, the r.m.s. radius, and the single-particle energies for the least-bound
neutron and proton states.
These results are obtained by recalculating the HF Hamiltonian using the rotated set of single-particle
states.
The description \textit{exact} in the figure captions refer to the numerical values when the
symmetry axis of the $^{20}$Ne nucleus is aligned with one of the code axes.
The results show the cumulative error for the $\beta$ rotation as is done from
$0^{\circ}$ to $45^{\circ}$. The spline order $M=3$ would be equivalent to using a three-point
formula for discretizing the derivative operators. As we can see, the results for higher-order
spline functions give reasonable accuracy for this rotation.
\begin{table}[!htb]
\caption{
Rotation with basis-spline order $M=9$ using the rotation method of Section~\ref{sec:direct}.
Here we are varying $\alpha$, $\beta$, and $\gamma$. The exact values for
the tabulated quantities are: $E_B=-157.26$~MeV, $r_{rms}=2.92$~fm, $\epsilon_{n}=-13.24$~MeV,
and $\epsilon_{p}=-9.35$~MeV.}
\label{tb:M3A}
\centering
\begin{tabular}{lrrrr}
\hline\hline
                & (45,45,45) & (90,90,90) & (135,135,135) & (180,180,180)\\
\hline
$E_B$           & -157.26 & -157.25 & -157.25 & -157.25\\
$r_{rms}$       &    2.92 &    2.92 &    2.92 &    2.92\\
$\epsilon_{n}$  &  -13.24 &  -13.24 &  -13.24 &  -13.24\\
$\epsilon_{p}$  &   -9.34 &   -9.34 &   -9.34 &   -9.34\\
\hline
\end{tabular}
\end{table}
In Table~\ref{tb:M3A} we tabulate the cumulative errors for rotating in all three Euler angles
for basis-spline order $M=9$. Again, while not perfect, the errors are reasonable
for this set of numerical parameters. The difference of the direct application method
discussed in this section relative to the ones discussed below is that the errors
accumulate due to the recursive nature of the computations.

\subsubsection{\label{sec:passive}Passive Rotation by Spline Interpolation}
In the BSCM approach, each component of the single-particle spinor is expanded in terms of basis-spline functions
of order $M$~\cite{US91}:
\begin{equation}
\psi(x,y,z)=\sum_{ijk} C_{ijk} B_{i}^{M}(x) B_{j}^{M}(y) B_{k}^{M}(z)\; ,
\label{eq:BSexp}
\end{equation}
where the $C_{ijk}$'s are expansion coefficients in terms of the basis-spline knots,
the $B$'s are basis-spline functions, and we have omitted the single-particle state
indices for notational simplicity.
In the basis-spline collocation method, $x$, $y$, and $z$ are then discretized on a
collocation lattice such that each component of the single-particle spinors attains the form
\begin{eqnarray*}
\psi(x_{\lambda},y_{\mu},z_{\nu})=\sum_{ijk} C_{ijk} B_{i}^{M}(x_{\lambda}) B_{j}^{M}(y_{\mu}) B_{k}^{M}(z_{\nu})\; ,
\end{eqnarray*}
which, for a given order $M$, may be expressed in matrix notation
\begin{equation}
\psi_{\lambda \mu \nu}=\sum_{ijk} C_{ijk} B_{i\lambda} B_{j\mu} B_{k\nu}\; ,
\label{eq:BSCexp}
\end{equation}
where $\lambda$, $\mu$, and $\nu$ are the indices for the three-dimensional Cartesian collocation lattice.
This expansion leads to the corresponding lattice representation of the Hartree-Fock equations~\cite{US91,UW91}.
After we obtain the collocation-lattice solutions of the Hartree-Fock equations in the
unrotated system, we can compute the expansion coefficients in Eq.~(\ref{eq:BSCexp}) as~\cite{UW91}
\begin{equation}
C_{ijk}=\sum_{\lambda \mu \nu} B^{\lambda i} B^{\mu j} B^{\nu k} \psi_{\lambda \mu \nu}\; ,
\label{eq:BScoeff}
\end{equation}
where $B^{\lambda i}$ denotes the inverse of $B_{i \lambda}$ and so on.
Consequently, the passive rotation operation given by Eq.~(\ref{eq:R8})
simply reduces to an interpolation of the single-particle states at the
rotated coordinate points
\begin{equation}
\psi(x',y',z')=\sum_{ijk} C_{ijk}B_{i}^{M}(x') B_{j}^{M}(y') B_{k}^{M}(z')\; .
\label{eq:BSexp2}
\end{equation}
We then rotate the resulting spinors in spin space via Eq.~(\ref{eq:Rs4}).
The expansion coefficients $C_{ijk}$ need only be calculated once for each
single-particle spinor component and can be used for as many $(\alpha, \beta, \gamma)$
combinations as are needed. The calculation of the expansion coefficients is the most
CPU-expensive part of this procedure and has the same dimensions as the wavefunction
array in the code. However, because we need to calculate the expansion coefficients
only once, the effective CPU time requirement
is comparable to or better than the direct active method mentioned earlier, depending on the
number of points used in the discretization of the Euler angles (more points is favorable
here).
\begin{table}[!htb]
\caption{
Rotation with basis-spline order $M=9$ using the rotation method of Section~\ref{sec:passive}.
Here we are varying $\alpha$, $\beta$, and $\gamma$. The exact values for
the tabulated quantities are: $E_B=-157.26$~MeV, $r_{rms}=2.92$~fm, $\epsilon_{n}=-13.24$~MeV,
and $\epsilon_{p}=-9.35$~MeV.}
\label{tb:M2A}
\centering
\begin{tabular}{lrrrr}
\hline\hline
               & (45,45,45) & (90,90,90) & (135,135,135) & (180,180,180)\\
\hline
$E_B$          & -157.25 & -157.26 & -157.25 & -157.26\\
$r_{rms}$      &    2.92 &    2.92 &    2.92 &    2.92\\
$\epsilon_{n}$ &  -13.24 &  -13.24 &  -13.24 &  -13.24\\
$\epsilon_{p}$ &   -9.34 &   -9.35 &   -9.34 &   -9.35\\
\hline
\end{tabular}
\end{table}

In Table~\ref{tb:M2A} we again tabulate the cumulative errors for rotating in all three Euler angles
for basis-spline order $M=9$. What is interesting for this case is that the errors do not
accumulate for larger angles because we are not reaching those angles recursively but by
direct interpolation. As a matter of fact, every $90^{\circ}$ rotation for each angle reproduces
the exact solutions due to the symmetry of the mesh in the three Cartesian directions. In this sense,
if an integral over the three Euler angles were considered (like in angular momentum projection)
the cumulative error is expected to be much smaller using this passive transformation.

\subsection{\label{sec:passive2}Rotation Before Variation} 
In a fully-unrestricted, three-dimensional calculation for a deformed nucleus, all orientations
of the nucleus are equivalent. In practical calculations the final orientation of the nucleus depends
on its initial orientation. In other words, the filling of the levels for the initial
guess for the HF single-particle states determines the final orientation. Normally this
guess results in a deformed nucleus whose symmetry axis (say an axially-symmetric nucleus) is
oriented along one of the principal axes of the code frame. However, by rotating the initial
state via the passive transformation given by Eq.~(\ref{eq:R8})and a subsequent rotation in spin-space,
we can obtain solutions to the HF equations that are oriented exactly at those angles with
respect to the unrotated solution. This method of \textit{rotation before variation} yields
the most accurate result of all three methods and can be trivially achieved with no interpolation.
The only discrepancies again seem to
arise for $45^{\circ}$ rotations, most likely because all the derivative operators are
generated along the principal axes of the code frame and may be slightly inefficient when
acting on a nucleus which is maximally-rotated with respect to the principal axes.
While this method may be more CPU-intensive if solutions are needed at many different
values of the Euler angles, it is very suitable for the initialization of time-dependent
Hartree-Fock (TDHF) calculations for deformed nuclei~\cite{UO06c,UO08a,UO10a,RU}. Since the initial
HF states are usually analytic, the computation of the states on the rotated coordinates
is trivial and the converged HF state is an eigenstate of the HF Hamiltonian without any
approximations, which is the most suitable initial state for TDHF calculations.
In Table~\ref{tb:M1A} we show the same cumulative errors shown for the previous two methods.
As we can see, this method is considerably more accurate for intermediate angles and is exact for
$90^{\circ}$ rotations for each angle.

We emphasize that this method will produce rotated states at the accuracy of the DFT code being
used and we find it to be the most useful procedure for TDDFT initialization of deformed nuclei.

\begin{table}[!htb]
\caption{
Rotation with basis-spline order $M=9$ using the rotation method of Section~\ref{sec:passive2}.
Here we are varying $\alpha$, $\beta$, and $\gamma$. The exact values for
the tabulated quantities are: $E_B=-157.26$~MeV, $r_{rms}=2.92$~fm, $\epsilon_{n}=-13.24$~MeV,
and $\epsilon_{p}=-9.35$~MeV.}
\label{tb:M1A}
\centering
\begin{tabular}{lrrrr}
\hline\hline
               & (45,45,45) & (90,90,90) & (135,135,135) & (180,180,180)\\
\hline
$E_B$          & -157.26 & -157.26 & -157.26 & -157.26 \\
$r_{rms}$      &    2.92 &    2.92 &    2.92 &    2.92\\
$\epsilon_{n}$ &  -13.24 &  -13.24 &  -13.24 &  -13.24\\
$\epsilon_{p}$ &   -9.35 &   -9.35 &   -9.35 &   -9.35\\
\hline
\end{tabular}
\end{table}

\section{Conclusions}
With the advances in computational power, it is increasingly possible to perform nuclear
structure and reaction calculations without any spatial symmetry assumptions.
One subset of such calculations involves the mean-field methods with the principal
wavefunction being a Slater determinant corresponding to a deformed nucleus.
In many situations, such as in collisions involving such nuclei, it is desirable to
study the reaction as a function of orientation of the deformed system.
In this work, we have presented
a comparative study of three methods to perform such rotations, the first being the standard active
rotation employed in many calculations. The other two methods involve a passive
rotation in coordinate space followed by an active rotation in spin space.
We have discussed the advantages and disadvantages of each method and have demonstrated
that the accurate implementation of three-dimensional rotations do require increased
accuracy in representing the differential operators on the lattice. The basis-spline
expansion provides one such approach, which has been demonstrated by the accuracy obtained for all
three methods. 

For three-dimensional TDDFT initialization of deformed nuclei with a particular alignment of their symmetry
axis with respect to the collision axis this task can be most easily achieved by 
constructing the initial guess for HF calculations (usually Cartesian oscillators) by evaluating it
on mesh values rotated with respect to the code axes. Subsequent HF iterations do not change this
orientation thus resulting in the desired HF solution. This procedure involves no interpolation
procedure and is the most straightforward method to implement in TDDFT codes.

\section*{Acknowledgments}
This work has been supported by the U.S. Department of Energy under grant No.
DE-FG02-96ER40975 with Vanderbilt University.

\end{document}